\documentclass[lettersize,conference]{IEEEtran}

\usepackage{amsmath,amsfonts}
\usepackage[ruled,linesnumbered]{algorithm2e}
\usepackage[caption=false,font=footnotesize,labelfont=rm,textfont=rm]{subfig}
\usepackage{makecell}
\usepackage{stfloats}
\usepackage{ragged2e}
\usepackage{url}
\usepackage{verbatim}
\usepackage{graphicx}
\usepackage{balance}
\usepackage{cite}
\usepackage{multirow}
\usepackage{enumitem}
\usepackage{bm}
\usepackage{subfig}
\usepackage{tikz}
\usetikzlibrary{shapes, arrows.meta, positioning}
\usepackage{booktabs}
\usepackage{lipsum}
\usepackage{amssymb}
\usepackage[colorlinks,linkcolor=blue,anchorcolor=blue,citecolor=blue]{hyperref}

\makeatletter
\patchcmd{\@makecaption}
{\\}
{.\ }
{}
{}

\newcommand{\myreffig}[1]{Fig. \ref{#1}}

\title{BeamCKMDiff: Beam-Aware Channel Knowledge Map Construction via Diffusion Transformer}

\author{
	Le Zhao$^1$, Yining Wang$^1$, Xinyi Wang$^1$, Zesong Fei$^{1}$, and Yong Zeng$^{2,3}$\\
	$^1$ School of Information and Electronics, Beijing Institute of Technology, Beijing, China\\
	$^2$ National Mobile Communications Research Laboratory, Southeast University, Nanjing, China\\
	$^3$ Purple Mountain Laboratories, Nanjing 211111, China\\
	Contact Email: tobin\_bit@icloud.com
}

\begin{document}
	\maketitle
	\begin{abstract}
		Channel knowledge map (CKM) is emerging as a critical enabler for environment-aware 6G networks, offering a site-specific database to significantly reduce pilot overhead. However, existing CKM construction methods typically rely on sparse sampling measurements and are restricted to either omnidirectional maps or discrete codebooks, hindering the exploitation of beamforming gain.
		To address these limitations, we propose BeamCKMDiff, a generative framework for constructing high-fidelity CKMs conditioned on arbitrary continuous beamforming vectors without site-specific sampling. Specifically, we incorporate a novel adaptive layer normalization (adaLN) mechanism into the noise prediction network of the Diffusion Transformer (DiT). This mechanism injects continuous beam embeddings as {global control parameters}, effectively steering the generative process to capture the complex coupling between beam patterns and environmental geometries.
		Simulation results demonstrate that BeamCKMDiff significantly outperforms state-of-the-art baselines, achieving superior reconstruction accuracy in capturing main lobes and side lobes.
		
	\end{abstract}
	
	
	\section{Introduction}
	\label{sec:intro}
	
	The evolution toward sixth-generation (6G) communication systems is driven by the demand for ubiquitous connectivity, ultra-high reliability, and massive capacity \cite{sunCKM6G, sun20256GAI}. Multi-input multi-output (MIMO) technology promises significant spectral efficiency gains but imposes severe challenges on beam training \cite{zeng2024tutorial}. In complex urban environments, the pilot overhead required for channel state information (CSI) feedback scales proportionally with the antenna array dimensions, becoming a bottleneck for system throughput. To mitigate this, the concept of channel knowledge map (CKM) has been proposed as a paradigm shift from pilot-centric to environment-aware communications \cite{Zeng_CKM_2021}. By providing location-specific channel parameters, CKM serves as a promising technique for efficient beam management and resource allocation.
	
	Early approaches to CKM construction primarily relied on interpolation techniques such as Kriging \cite{sato2017kriging}. While computationally inexpensive, these methods assume spatial stationarity and fail to model the abrupt signal attenuation caused by shadowing in dense urban canyons. Furthermore, high-fidelity CKM construction typically demands extensive sampling measurements, which is often impractical. Alternatively, model-driven methods like ray-tracing offer high physical accuracy through deterministic modeling. However, their excessive computational complexity renders them unsuitable for real-time or large-scale CKM construction.
	
	The integration of artificial intelligence (AI) has revolutionized CKM construction. Pioneering works like RadioUNet \cite{levie2021radiounet} and IMNet \cite{le2025IMNet} utilized convolutional neural networks to learn electromagnetic propagation features, treating map construction as an image translation task. Subsequent studies further explored various neural architectures to improve accuracy, including graph neural networks \cite{chen2023graph}, graph attention networks \cite{li2023graph}, and generative adversarial networks \cite{Zhang_RMEGAN_2023}. Recently, denoising diffusion probabilistic models (DDPMs) have demonstrated superior capability in modeling complex electromagnetic characteristics. For instance, RadioDiff \cite{wang2024radiodiff, le20253Dradiodiff} applied DDPMs to construct CKMs, successfully recovering high-frequency texture details and sharp shadow boundaries that regression-based methods often smooth out. Experimental results consistently indicate that diffusion-based CKM construction yields superior performance compared to traditional baselines.
	
	Note that the aforementioned works primarily focused on constructing omni-directional CKMs for single-antenna systems, while modern communication systems are extensively equipped with multi-antenna arrays, necessitating beam-aware CKM capabilities. Recently, \cite{Wang_BeamCKM_2025} proposed BeamCKM to address this by constructing CKMs for specific codebooks using an attention-based architecture. However, BeamCKM treats beam indices as categorical embeddings rather than continuous vectors. This approach fundamentally fails to capture the continuous relationship between beamforming vectors and spatial energy distribution, limiting the model to a finite, predefined set of discrete patterns, which inevitably introduces beam quantization degradation when users are misaligned with pre-defined grid directions.
	
	To bridge this gap, we propose BeamCKMDiff, a generative framework for constructing high-fidelity CKMs conditioned on arbitrary continuous beamforming vectors. \textit{To the best of our knowledge, this is the first work to address CKM construction in the continuous beam domain.} Specifically, leveraging a diffusion transformer (DiT), we introduce a conditional generation mechanism where continuous beamforming vectors are encoded as {global control parameters}. By injecting these descriptors via adaptive layer normalization (adaLN), the model dynamically aligns the generated CKMs with the physical beam gain distribution under the environmental conditions. Simulations on high-fidelity ray-tracing datasets demonstrate that BeamCKMDiff achieves state-of-the-art performance.
	
	The remainder of this paper is organized as follows. Section \ref{sec:system} describes the system model and problem formulation. Section \ref{sec:method} details the proposed BeamCKMDiff framework. Section \ref{sec:simulation} presents the experimental setup and simulation results. Finally, Section \ref{sec:conclusion} concludes the paper.
	
	\section{System Model and Problem Formulation}
	\label{sec:system}
	\subsection{System Model}
	We consider a downlink multi-antenna communication system deployed in a dense urban region $\mathcal{D}$ of physical dimensions $\{H \times W\}$ meters. The region is discretized into a uniform grid map of size $\{H/\Delta \kappa \times W/\Delta \kappa\}$ pixels, where $\Delta \kappa$ denotes the resolution. Each pixel $(x, y)$ represents a specific physical location $\mathbf{q} = [x,y]^T \in \mathbb{R}^2$. The environment is characterized by a static building distribution map $\mathbf{B} \in \mathbb{R}^{H \times W}$. Let $\mathcal{B} \subset \mathcal{D}$ denote the set of locations occupied by buildings. The value ${B}_{x,y}$ represents the building height at $[x,y]$. The set of locations occupied by buildings is denoted as $\mathcal{B} \subset \mathcal{D}$, and the target CKM covers the free space $\mathcal{D} \setminus \mathcal{B}$. A ground base station (GBS) is deployed at location $\mathbf{q}_{\rm tx} \in \mathcal{D}$ with a specific height, equipped with a reconfigurable antenna array consisting of $N_t$ antennas.
	
	Given an arbitrary precoding vector $\mathbf{w} \in \mathbb{C}^{N_t}$, the transmitted signal vector is given by $\mathbf{x} = \mathbf{w} s$, where $s$ is the data symbol with normalized power $\mathbb{E}[|s|^2]=1$, and the beamforming vector satisfies the power constraint $\|\mathbf{w}\|^2 \le P_{max}$.
	The received signal $y(\mathbf{q})$ at the location is expressed as
	\begin{equation}
		y(\mathbf{q}) = \mathbf{h}^H(\mathbf{q}) \mathbf{w} s + n,
	\end{equation}
	where $\mathbf{h}(\mathbf{q}) \in \mathbb{C}^{N_t \times 1}$ represents the channel vector capturing the propagation effects, including path loss, shadowing, and multipath fading, and $n \sim \mathcal{CN}(0, \sigma^2)$ is the additive white Gaussian noise.
	The average received signal strength (RSS) can be calculated as
	\begin{align}
		\mathbb{E}\left[\left\vert y(\mathbf{q}) \right\vert^2\right] = \left\vert \mathbf{h}^H(\mathbf{q}) \mathbf{w} \right\vert^2 + \sigma^2.
	\end{align}
	For a specific beamforming vector $\mathbf{w}$, the beam-aware CKM, denoted as $\mathbf{\Psi}_{\mathbf{w}} \in \mathbb{R}^{H \times W}$, is defined as
	\begin{equation}
		{\Psi}_{\mathbf{w}}(\mathbf{q}) = 10 \log_{10} \left( \left\vert \mathbf{h}^H(\mathbf{q}) \mathbf{w} \right\vert^2 \right) \in \mathbf{\Psi}_{\mathbf{w}}, \quad \forall \mathbf{q} \in \mathcal{D} \setminus \mathcal{B}.
	\end{equation}
	This formulation explicitly shows that the map $\mathbf{\Psi}_{\mathbf{w}}$ is a function of both the environment (implicit in $\mathbf{h}$) and the continuous control variable $\mathbf{w}$.

	Crucially, this continuous parameterization is designed to encode the beamforming weights—specifically the complex magnitude and phase adjustments applied to each antenna element—into the spatial mapping process. This enables the model to support flexible synthesis of beamforming vectors for arbitrary directions beyond finite codebooks, significantly improving practicality and adaptability for dynamic multi-antenna systems.
	
	\subsection{Problem Formulation}
	Based on the system model described above, we frame the CKM construction as a conditional generative modeling problem.  Our objective is to learn the conditional probability distribution $p(\mathbf{\Psi}_{\mathbf{w}} | \mathbf{C})$, where the condition set is defined as $\mathbf{C} = \{ \mathbf{B}, \mathbf{T}, \mathbf{w} \}$. Here, $\mathbf{T} \in \{0,1\}^{H/\Delta \kappa \times W/\Delta \kappa}$ is the sparse transmitter location map, where $T_{\mathbf{q}}$ is set as for $\mathbf{q} = \mathbf{q}_{\rm tx}$, and 0 otherwise.
	
	Formally, we aim to approximate the mapping function $\mathcal{F}_{\theta}$ parameterized by a neural network expressed as
	\begin{equation}
		\hat{\mathbf{\Psi}}_{\mathbf{w}} = \mathcal{F}_{\theta}(\mathbf{B}, \mathbf{T}, \mathbf{w}),
	\end{equation}
	such that the generated map $\hat{\mathbf{\Psi}}_{\mathbf{w}}$ minimizes the reconstruction error $\|\hat{\mathbf{\Psi}}_{\mathbf{w}} - \mathbf{\Psi}_{\mathbf{w}}\|^2$ across the entire spatial domain {while faithfully reproducing the complex fading patterns, reflections, and shadowing effects characteristic of wireless signal propagation.}
	Through iteratively training $\mathcal{F}_{\theta}$, the model learns to reconstruct the high-fidelity ground-truth $\mathbf{\Psi}_{\mathbf{w}}$ for any arbitrary continuous beamforming vector $\mathbf{w}$.

	\section{Proposed BeamCKMDiff}
	\label{sec:method}
\begin{figure}[!t]
		\centering
		\includegraphics[width=1\linewidth]{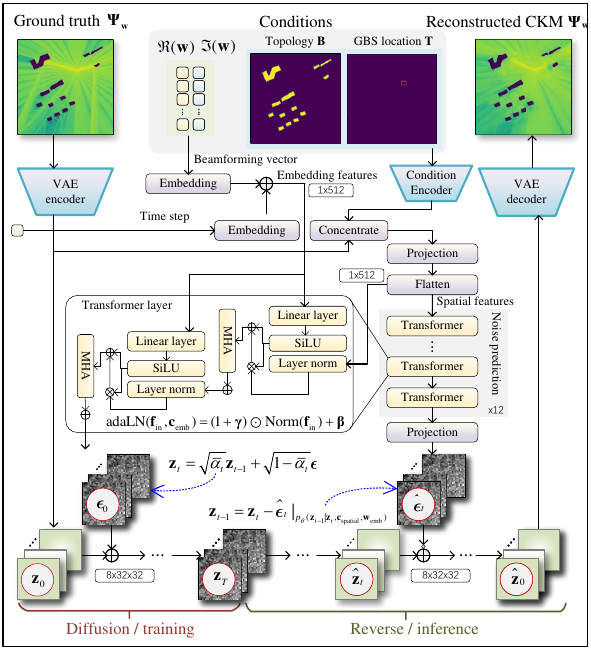}
		\caption{The illustration of the proposed BeamCKMDiff. The VAE is employed to encode the CKM into latent space, thereby reducing the dimension of the diffusion model.
		}
		\label{fig:BeamCKMDiff_framework}
		%		\vspace*{-0.2cm}
	\end{figure}

	In this section, we present the proposed BeamCKMDiff architecture. As illustrated in Fig. \ref{fig:BeamCKMDiff_framework}, the proposed framework consists of a variational autoencoder (VAE) for perceptual compression, a condition encoder for environment topology spatial feature extraction, and a DiT-based backbone with beamforming vector embedding for beam-aware CKM generative reconstruction.
	%	\vspace{-0.3cm}
	
	\vspace{-0.2cm}
	\subsection{Perceptual Compression via VAE}
	Directly training diffusion models in the high-resolution pixel space is computationally extensive and unstable. To mitigate this, we employ a VAE to project the high-dimensional CKM into a compact latent space.
	
	The VAE consists of an encoder $\mathcal{E}(\cdot)$ and a decoder $\mathcal{D}(\cdot)$. Given a ground-truth CKM $\mathbf{\Psi}_{\mathbf{w}}$, the encoder maps it to a latent distribution expressed as $q(\mathbf{z}|\mathbf{\Psi}_{\mathbf{w}}) = \mathcal{N}(\mathbf{z}; \bm{\mu}, \bm{\sigma}^2)$, where $\mathbf{z} \in \mathbb{R}^{h \times w \times D_{\rm lat}}$ with $h = H/f$ and $w = W/f$. The decoder aims to reconstruct the original map from the sampled latent $\mathbf{z}$.

	Prior to training the condition encoder and DiT-based denoising module, we pre-train the VAE.
	The training objective is to minimize the evidence lower bound (ELBO), combining a reconstruction loss and a KL-divergence regularization:
	\begin{equation}
		\mathcal{L}_{\rm VAE} = \|\mathbf{\Psi}_{\mathbf{w}}- \mathcal{D}(\mathbf{z})\|_2^2 + \lambda_{\rm KL} D_{\rm KL}(q(\mathbf{z}|\mathbf{\Psi}_{\mathbf{w}}) \| \mathcal{N}(\mathbf{0}, \mathbf{I})).
	\end{equation}
	Once trained, the VAE is frozen, and the diffusion process operates entirely within the efficient latent space.
	
	%	\vspace{-0.2cm}
	\subsection{Condition Encoder}
	\label{subsec:condition_encoder}
	To effectively extract semantic features that are spatially aligned with the latent noise, our condition encoder employs a deep ResNet-based architecture to extract semantic features that are spatially aligned with the latent noise.
	The input to the condition encoder is a multi-channel tensor with dimensions of $\mathbb{R}^{H \times W \times 2}$, constructed by concatenating the environment topology $\mathbf{B}$ and GBS location map $\mathbf{T}$.
	The encoder applies a series of convolution modules and residual blocks to downsample the condition inputs to match the spatial resolution of the latent space with dimension. The output is a spatial condition embedding as expressed as
	\begin{align}\label{enc_token}
		\mathbf{c}_{\rm env} = {\rm Proj}\left(\rm{Concat}(\mathbf{B}, \mathbf{T})\right)\in \mathbb{R}^{h \times w \times D_{\rm cond}},
	\end{align}
	which encodes physical constraints such as {environmental layout information and transmitter-receiver spatial relationships}. The condition encoder is deterministic and trained end-to-end with the DiT, allowing it to adaptively learn optimal feature representations for the generative CKM reconstruction.
	
	\subsection{Continuous Beam Conditioning via DiT}
	DiT backbone \cite{peebles2023scalable} is the core of BeamCKMDiff. We treat the latent variable $\mathbf{z}_t$ at diffusion step $t$ and the spatial condition $\mathbf{e}_{\rm env}$ as sequences of tokens. The diffusion step $t$ and the continuous beamforming vector $\mathbf{w}$ are embedded into feature vectors, which are then used to generate the parameters for adaLN mechanism. The main components are shown in the following parts.
	
	% {This provides global modulation of the noise prediction network, conditioning it on the specific beam configuration and diffusion process.}
	
	% The diffusion step $t$ and beamforming vector $\mathbf{w}$ are then embedded into feature tokens to optimize the weights parameters of adaLN for global modulation. The main components are shown in the following parts.
	
	\subsubsection{Latent Feature Fusion}
	The noisy latent $\mathbf{z}_t \in \mathbb{R}^{h \times w \times D_{\rm lant}}$ and the spatial condition $\mathbf{e}_{\rm env}$ are concatenated to integrate the spatial environmental guidance, which can be expressed as
	\begin{align}
		\mathbf{c}_{\rm spatial} = {\rm Concat}(\mathbf{z}_t, \mathbf{c}_{\rm env}) \in \mathbb{R}^{h \times w \times (D_{\rm lat} + D_{\rm cond})}.
	\end{align}
	This fused map is then projected into a sequence of tokens, which are linearly embedded and augmented with learnable positional embeddings before entering the Transformer blocks.
	
	\subsubsection{Beam-Aware adaLN}
	Inspired by the class-conditional embedding strategy employed in DiT, the proposed Beam-CKMDiff framework introduces a novel conditioning mechanism that injects the continuous beamforming vector $\mathbf{w}$ into the diffusion process via adaLN. This mechanism dynamically modulates the channel statistics of the latent feature maps, thereby steering the generative process to align with specific beamforming configurations.
	
	Specifically, we fuse the time-step embedding $\mathbf{t}_{\rm emb}$ and the beamforming vector embedding $\mathbf{w}_{\rm emb}$ to obtain a global conditioning token expressed as $\mathbf{c}_{\rm emb} = \mathbf{t}_{\rm emb} + \mathbf{w}_{\rm emb}$.
	This token serves as a {global control parameters} containing both diffusion progress and physical beamforming information.
	Inside each DiT block, we employ a linear projection layer following a SiLU activation to dynamically generate the modulation parameters from $\mathbf{c}_{\rm emb}$. The scale factor $\bm{\gamma}$ and shift factor $\bm{\beta}$ are derived as
	\begin{equation}
		[\bm{\gamma}, \bm{\beta}] = \mathbf{A}_{\rm mod} \cdot \text{SiLU}(\mathbf{c}_{\rm emb}) + \mathbf{b}_{\rm mod},
	\end{equation}
	where $\mathbf{A}_{\rm mod}$ and $\mathbf{b}_{\rm mod}$ are learnable weights and biases of the linear layers in the adaLN module. Physically, $\bm{\gamma}$ acts as a gain controller, adjusting the amplitude or difference of the spatial features to highlight high-energy beam lobes, while $\bm{\beta}$ acts as a bias controller, shifting the baseline signal level to match the ground truth RSS distribution.
	{The adaLN is then outputting the latent tokens to multi-head attention (MHA) layers, by modulating the features $\mathbf{c}_{\rm spatial}$ expressed as}
	\begin{equation}
		\text{adaLN}(\mathbf{f}_{\rm in}, \mathbf{c}_{\rm emb}) = (\mathbf{1} + \bm{\gamma}) \odot \mathcal{N}(\mathbf{f}_{\rm in}) + \bm{\beta},
	\end{equation}
	where $\odot$ denotes element-wise multiplication, $\mathcal{N}(\cdot)$ is the normalization procedure. By learning to predict optimal $\bm{\gamma}$ and $\bm{\beta}$, the model steers the mean and variance of the intermediate features to align the generated CKM with the directionality and beam pattern of $\mathbf{w}$.

	\subsection{Diffusion Training and Interference}
	\subsubsection{Training / Diffusion Process}
	During training, we sample a random time step $t \sim \mathcal{U}(0, T)$, a Gaussian noise term $\bm{\epsilon} \sim \mathcal{N}(\mathbf{0}, \mathbf{I})$, and a training tuple $\{\mathbf{z}_0, \mathbf{c}_{\text{spatial}}, \mathbf{w}\}$. The forward diffusion process progressively corrupts the clean latent $\mathbf{z}_0$ into a noisy state $\mathbf{z}_t$. Leveraging the property of Gaussian transitions, $\mathbf{z}_t$ can be sampled directly via the closed-form expression:
	\begin{equation}
		\mathbf{z}_t = \sqrt{\bar{\alpha}_t} \mathbf{z}_0 + \sqrt{1 - \bar{\alpha}_t} \bm{\epsilon},
	\end{equation}
	where $\bar{\alpha}_t = \prod_{s=1}^t \alpha_s = \prod_{s=1}^t(1- \beta_s)$ determines the noise schedule, and $\beta_t, t\in \{1,\dots, T\}$ is a set of pre-defined parameters.
	
	The DiT model $\bm{\epsilon}_\theta$ takes the noisy latent $\mathbf{z}_t$, the time step $t$, and the condition set $\{\mathbf{c}_{\text{spatial}}, \mathbf{w}_{\rm emb}$\} as inputs to predict the added noise. It is trained to approximate the score function of the data distribution conditioned on $\mathbf{c}_{\text{spatial}}$ and $\mathbf{w}_{\rm emb}$. The network parameters $\theta$ are trained to minimize the simplified mean squared error (MSE) loss:
	\begin{equation}
		\mathcal{L}_{\text{diff}} = \mathbb{E}_{t, \mathbf{z}_0, \bm{\epsilon}, \mathbf{w}} \left[ \| \bm{\epsilon} - \bm{\epsilon}_\theta(\mathbf{z}_t, \mathbf{c}_{\text{spatial}}, \mathbf{w}_{\rm emb}, t) \|^2 \right], 
	\end{equation}
	By minimizing this objective, the model learns to reverse the diffusion process with the physical conditions provided by $\mathbf{c}_{\text{spatial}}$ and the beamforming guidance from $\mathbf{w}_{\rm emb}$.
	
	\subsubsection{Inference / Reverse Process}
	During inference, the generation process begins with pure Gaussian noise $\mathbf{z}_T \sim \mathcal{N}(\mathbf{0}, \mathbf{I})$. The model iteratively denoises the latent variable from $t=T$ to $0$ via the ancestral sampling procedure. At each step $t$, the model's noise prediction $\bm{\epsilon}_\theta$ is used to parameterize the reverse conditional distribution $p_\theta(\mathbf{z}_{t-1} | \mathbf{z}_t, \mathbf{c}_{\text{spatial}}, \mathbf{w}_{\rm emb})$. Sampling from this distribution yields the updated latent variable shown as
	\begin{align}
		\mathbf{z}_{t-1}=\mathbf{z}_t - \hat{\bm{\epsilon}}_t,
	\end{align}
	where the estimated noise $\hat{\bm{\epsilon}}_t$ is computed as a scaled combination of the model prediction and additional stochastic noise expressed as
	\begin{equation}
		\hat{\bm{\epsilon}}_t = \frac{1}{\sqrt{\alpha_t}} \left( \mathbf{z}_t - \frac{1-\alpha_t}{\sqrt{1-\bar{\alpha}_t}} \bm{\epsilon}_\theta(\mathbf{z}_t, \mathbf{c}_{\text{spatial}}, \mathbf{w}_{\rm emb}, t)\right) + \sigma_t \mathbf{z},\\
	\end{equation}
	where $\mathbf{z} \sim \mathcal{N}(\mathbf{0}, \mathbf{I})$ represents the stochastic noise injection for $t > 0$. Finally, the cleaned latent $\mathbf{z}_0$ is passed through the VAE decoder to reconstruct the high-fidelity beam-specific CKM $\hat{\mathbf{\Psi}} = \mathcal{D}(\mathbf{z}_0)$.

	\section{Simulation Results}
	\label{sec:simulation}
	This section evaluates the performance of the proposed BeamCKMDiff framework. We begin by introducing the simulation setup, including the dataset and parameter configurations. Subsequently, we compare our BeamCKMDiff against SOTA methods to assess their accuracy in Beam-aware CKM construction.
	%	\vspace{-0.3cm}
	
	\subsection{Simulation Settings}
	
	The BeamCKM dataset is constructed from 30 geo-referenced urban topologies extracted from OpenStreetMap \cite{osm_planetdump_2017}. Each scenario covers a 512 m $\times$ 512 m area, discretized into a grid with a spatial resolution of $2$ m. For each map, a Ground Base Station (GBS) equipped with a $16 \times 1$ ULA is deployed at 10 random locations with a height of 1.5 m.
	For each deployment, 10 random beamforming vectors are applied to generate diverse propagation samples. High-fidelity channel gain is generated using the NVIDIA Sionna ray-tracing engine \cite{hoydis2022sionna}. The sionna solver is configured with $10^9$ rays, a maximum of 3 reflections, and enabled diffraction. The system operates at a carrier frequency of $2.4$ GHz, and electromagnetic material properties follow ITU-R standards.
	
	During the forward diffusion process, noise is injected into the latent representation over $T=500$ timesteps, governed by a linear variance schedule ranging from $\beta_1 = 4\times 10^{-5}$ to $\beta_T = 5\times 10^{-3}$. The model parameters are optimized using the Adam optimizer with a learning rate of $10^{-4}$. The complete architecture and specific hyperparameter configurations for the VAE and BeamDiT modules are detailed in Table \ref{table_model_summary}. All training and inference procedures were executed on a workstation equipped with a single NVIDIA GeForce RTX 4090 GPU.

	\begin{table}[!t]
		\renewcommand{\arraystretch}{1.15} % 稍微增加行高以提高可读性
		\centering
		\setlength{\tabcolsep}{1.5mm}
		\caption{Model Architecture Configurations}
		\label{table_model_summary}
		\begin{tabular}{l|l}
			\Xhline{1.1pt}
			\textbf{Module} & \textbf{Core Components \& Hyperparameters} \\ \hline
			\textbf{VAE} & \textit{Input}: $1 \times 256 \times 256$, \textit{Latent}: $8 \times 32 \times 32$ \\
			\quad - \textit{Encoder} & Input $\xrightarrow{\text{Conv}}$ [Downsample $\to$ ResNetBlock]$_{\times 3}$ \\
			& Channels: $1 \to [64 \to 128 \to 256] \to 8$ \\
			& \texttt{map\_latent}: Conv2d(256, 8, $3\times3$) \\
			\quad - \textit{Decoder} & Input $\mathbf{z}$ $\xrightarrow{\text{Conv}}$ [ResNetBlock $\to$ Upsample $\to$ Conv]$_{\times 3}$ \\
			& Channels: $8 \to 256 \to [128 \to 64 \to 32] \to 1$ \\
			& \texttt{final}: Conv2d(32, 1, $3\times3$) $\to$ Sigmoid \\ \hline
			\textbf{ResNetBlock} & \texttt{gn1}: GroupNorm(8, $C_{in}$) $\to$ SiLU \\
			& \texttt{conv1}: Conv2d($C_{in}, C_{out}, 3\times3$) \\
			& \texttt{gn2}: GroupNorm(8, $C_{out}$) $\to$ SiLU \\
			& \texttt{conv2}: Conv2d($C_{out}, C_{out}, 3\times3$) \\ \hline
			\textbf{Cond. Encoder} & \textit{Input}: $3 \times 256 \times 256$, \textit{Output}: $32 \times 32 \times 32$ \\
			& \texttt{head}: Conv2d(3, 32, $3\times3$) \\
			& Body: [ResNetBlock $\to$ Downsample]$_{\times 3}$ \\
			& Channels: $32 \to [32 \to 64 \to 128]$ \\
			& \texttt{out}: Zero-init Conv2d(128, 32, $3\times3$) \\ \hline
			\textbf{BeamDiT} & \textit{Input}: $40 \times 32 \times 32$ ($8_{\text{lat}} + 32_{\text{cond}}$) \\
			\quad - \textit{Embedders} & \texttt{x\_emb}: Conv2d(40, 512, $2\times2$, stride=2) \\
			& \texttt{pos\_emb}: Learnable ($1 \times 256 \times 512$) \\
			& \texttt{t/w\_emb}: MLP(Linear $\to$ SiLU $\to$ Linear), $D=512$ \\
			\quad - \textit{Backbone} & \textbf{Depth}: 12, \textbf{Heads}: 8, \textbf{Hidden}: 512 \\
			\quad - \textit{DiT Block} & \texttt{adaLN}: Regresses params ($\gamma, \beta$) \\
			& \texttt{attn}: MultiheadAttention (dim=512, heads=8) \\
			& \texttt{mlp}: Linear(512 $\to$ 2048) $\to$ GELU $\to$ Linear(512) \\
			\quad - \textit{Final Layer} & \texttt{norm}: LayerNorm + adaLN (Scale $\gamma$, Shift $\beta$) \\
			& \texttt{linear}: Linear(512, $2^2 \times 8$) \\
			\Xhline{1.1pt}
		\end{tabular}
		%		\vspace{-0.1cm}
	\end{table}
	
	To evaluate the performance of the proposed framework, we compare BeamCKMDiff against three representative methods, each adopting a different modeling paradigm:
	
	\begin{enumerate}
		\item {RadioUNet \cite{levie2021radiounet}:} A classic discriminative model based on the U-Net architecture. It learns a deterministic mapping from environmental geometries to CKMs using a UNet structure with skip connections.
		\item {TransUNet \cite{Wang_BeamCKM_2025}:} A transformer-UNet based  framework designed for CKM. It incorporates beam information by embedding discrete beam indices from a DFT codebook into the network, handling beamforming as a categorical condition rather than a continuous vector.
		\item {RadioDiff-UNet \cite{wang2024radiodiff}:} A baseline diffusion probabilistic model that employs a standard U-Net backbone. It relies on convolutional layers for denoising and does not utilize the adaptive layer normalization (adaLN) mechanism for continuous global conditioning.
	\end{enumerate}
	
	We adopt the Normalized MSE (NMSE) in dB to evaluate the quality of the constructed CKM, which is expressed as
	\begin{equation}
		\label{nmse_define}
		\text{NMSE (dB)} = 10 \log_{10} \left( \frac{\sum_{\mathbf{q} \in \mathcal{D/B}} | \widehat{\Psi}(\mathbf{q}) - \Psi(\mathbf{q}) |^2}{\sum_{\mathbf{q} \in \mathcal{D/B}} | \Psi(\mathbf{q}) |^2} \right),
	\end{equation}
	where $\Psi(\mathbf{p})$ and $\widehat{\Psi}(\mathbf{q})$ denote the ground truth and predicted pathloss values at grid location $\mathbf{q}$, respectively.
	
	To evaluate model generalization, we established two scenarios:
	\begin{itemize}
		\item \textbf{Scenario 1 (Unseen Beams)}: Testing with trained locations but conditioned on new continuous beamforming vectors to assess beam manifold interpolation.
		\item \textbf{Scenario 2 (Unseen Locations)}: Testing with new transmitter locations to assess spatial interpolation capabilities.
		%		{\item \textbf{Scenario 3 (Unseen Topology)}: Testing on unseen environments with random transmitter locations and beamforming vectors to assess spatial interpolation capabilities.}
	\end{itemize}
	%	\vspace*{-0.2cm}
	
	\begin{figure*}[!t]
		\centering
		\subfloat{\includegraphics[width=1.1in]{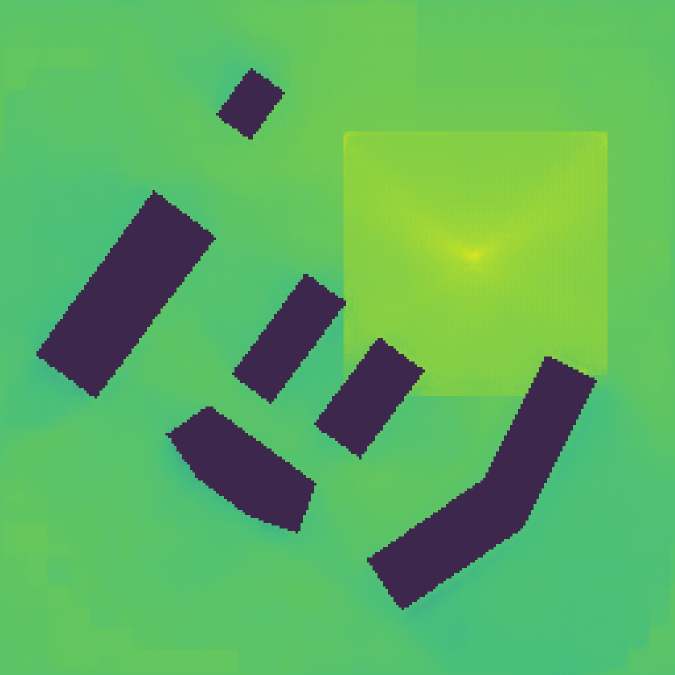}}
		\hspace{0.25cm}
		\subfloat{\includegraphics[width=1.1in]{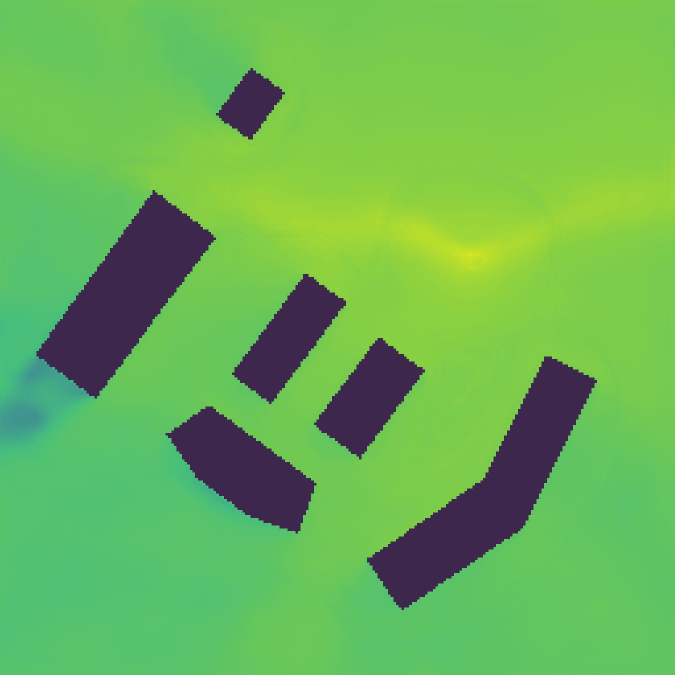}}
		\hspace{0.25cm}
		\subfloat{\includegraphics[width=1.1in]{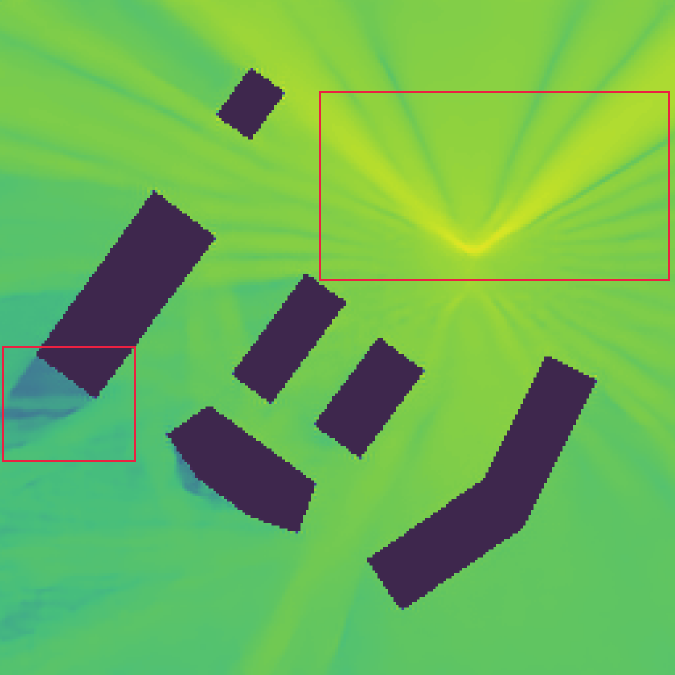}}
		\hspace{0.25cm}
		\subfloat{\includegraphics[width=1.1in]{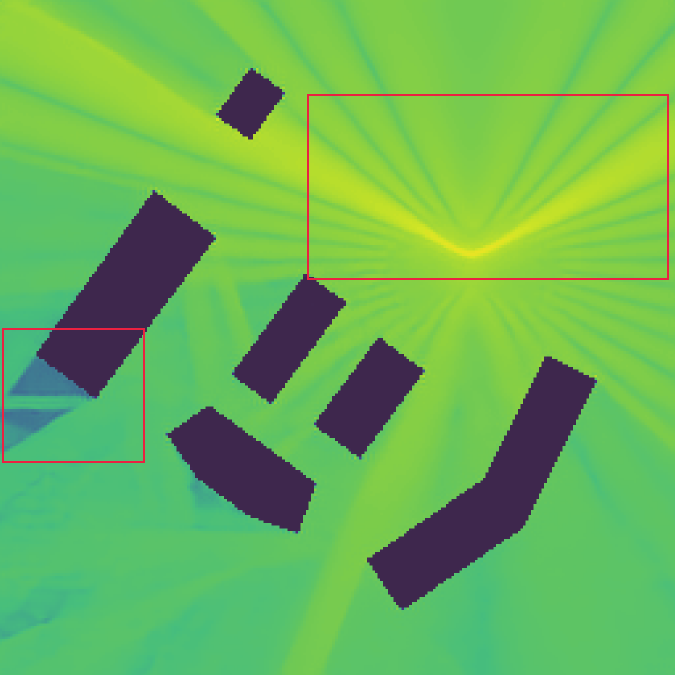}}
		\hspace{0.25cm}
		\subfloat{\includegraphics[width=1.1in]{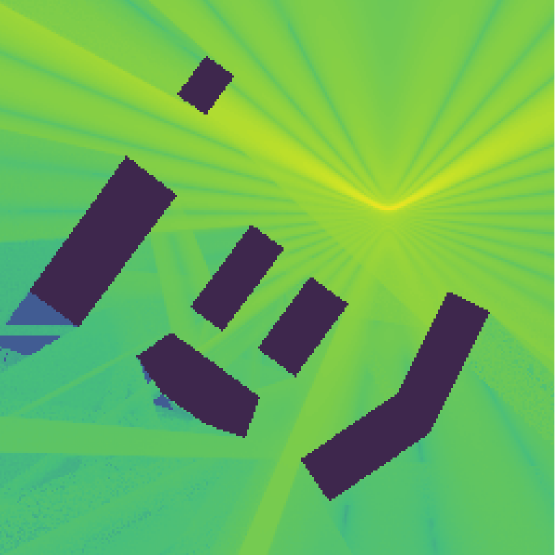}}\\
		\setcounter{subfigure}{0}
		\subfloat[RadioUNet]{\includegraphics[width=1.1in]{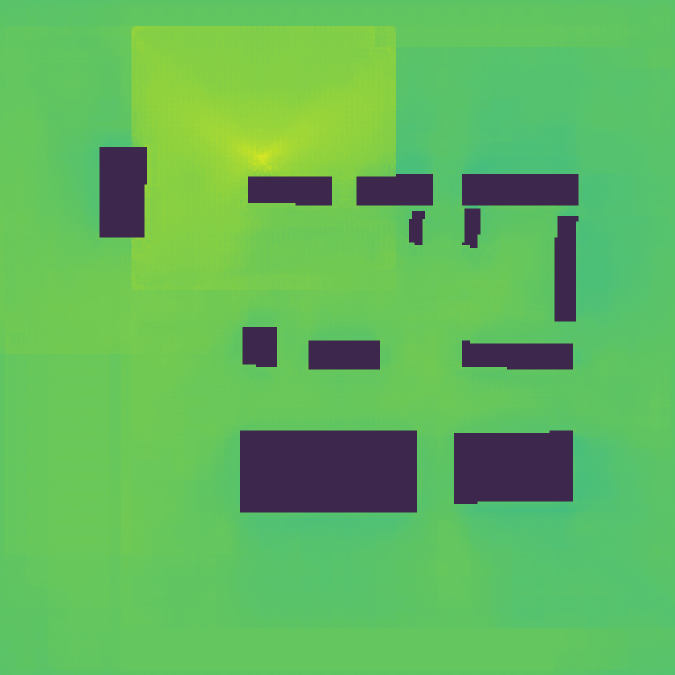}}
		\hspace{0.25cm}
		\subfloat[TransUNet]{\includegraphics[width=1.1in]{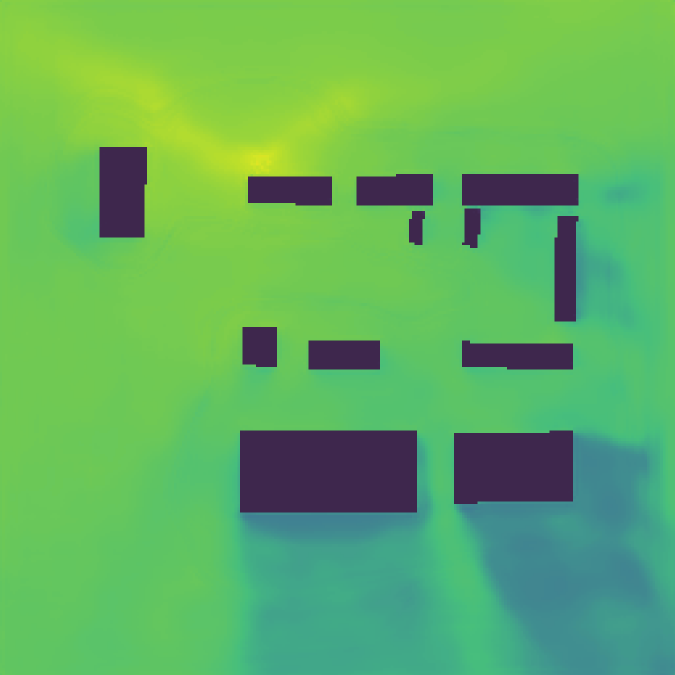}}
		\hspace{0.25cm}
		\subfloat[RadioDiff-UNet]{\includegraphics[width=1.1in]{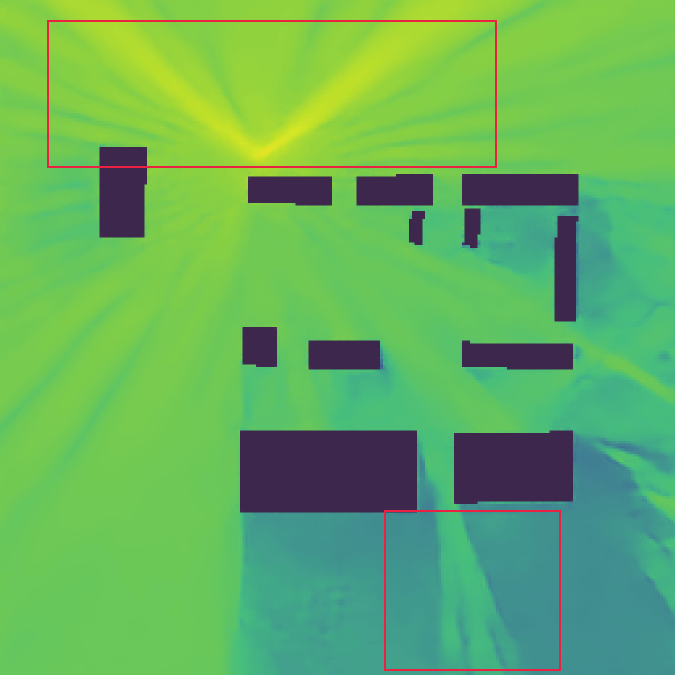}}
		\hspace{0.25cm}
		\subfloat[BeamCKMDiff]{\includegraphics[width=1.1in]{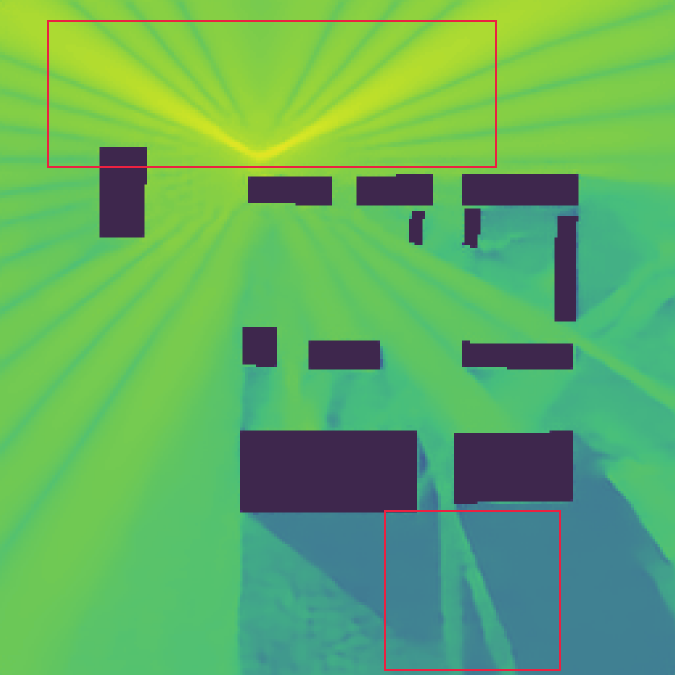}}
		\hspace{0.25cm}
		\subfloat[Ground Truth]{\includegraphics[width=1.1in]{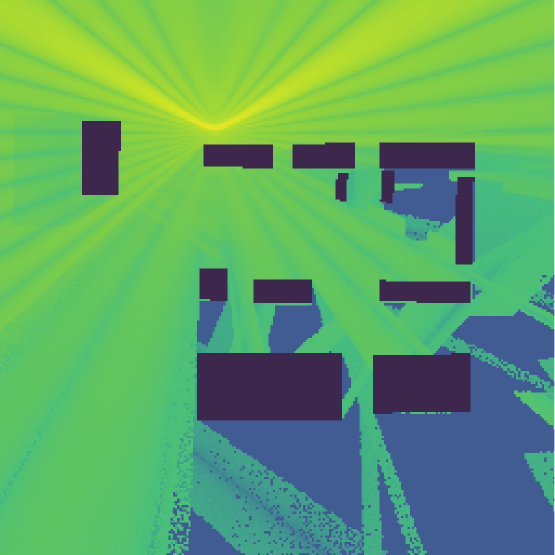}}
		\caption{Comparisons of constructed CKM via different methods under \textbf{seen GBS locations} and \textbf{unseen beams}.}
		\label{fig:visualization_umseemed_beam}
		%		\vspace{-0.4cm}
	\end{figure*}
	
	\begin{figure*}[!t]
		\centering
		\subfloat{\includegraphics[width=1.1in]{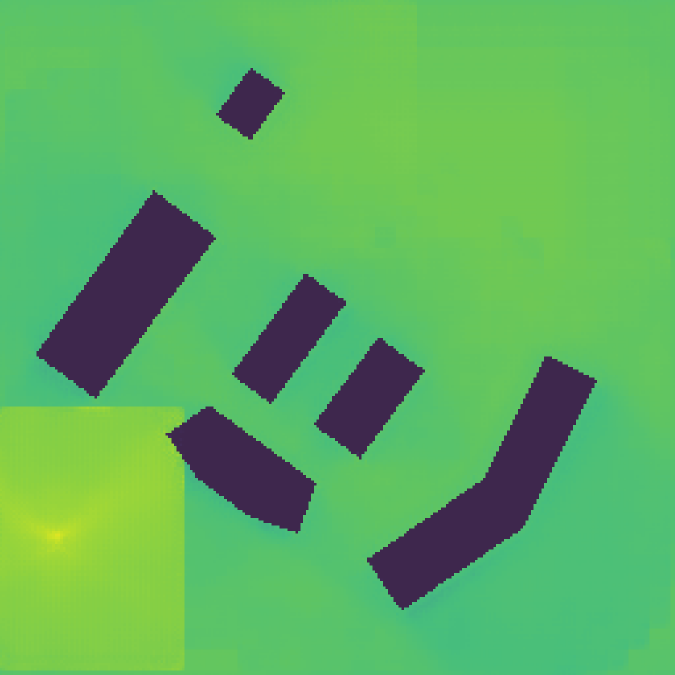}}
		\hspace{0.25cm}
		\subfloat{\includegraphics[width=1.1in]{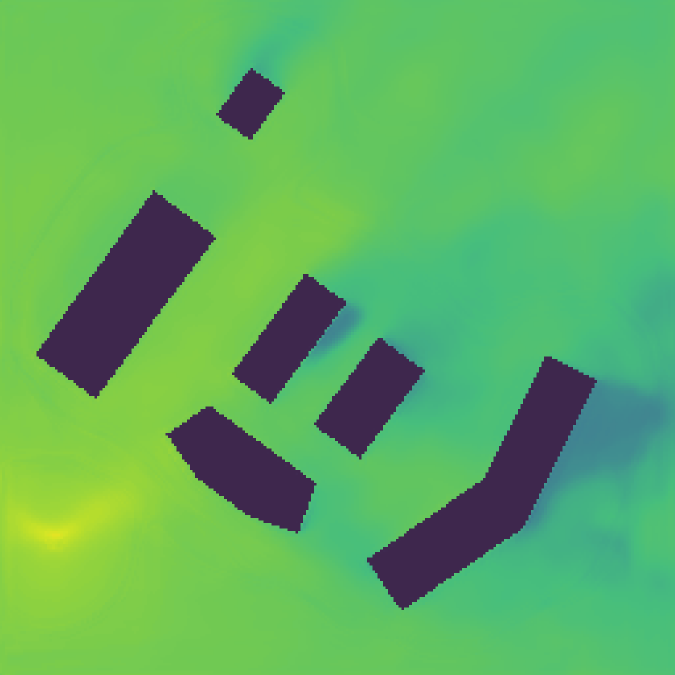}}
		\hspace{0.25cm}
		\subfloat{\includegraphics[width=1.1in]{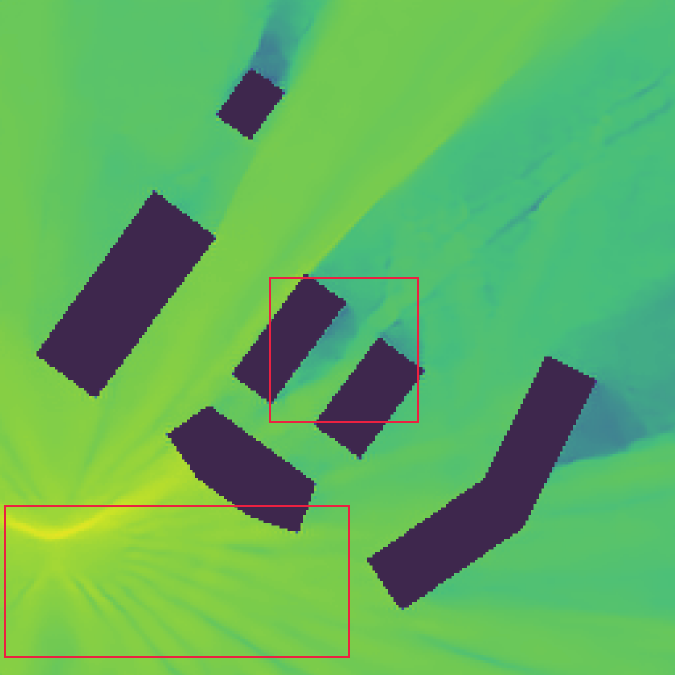}}
		\hspace{0.25cm}
		\subfloat{\includegraphics[width=1.1in]{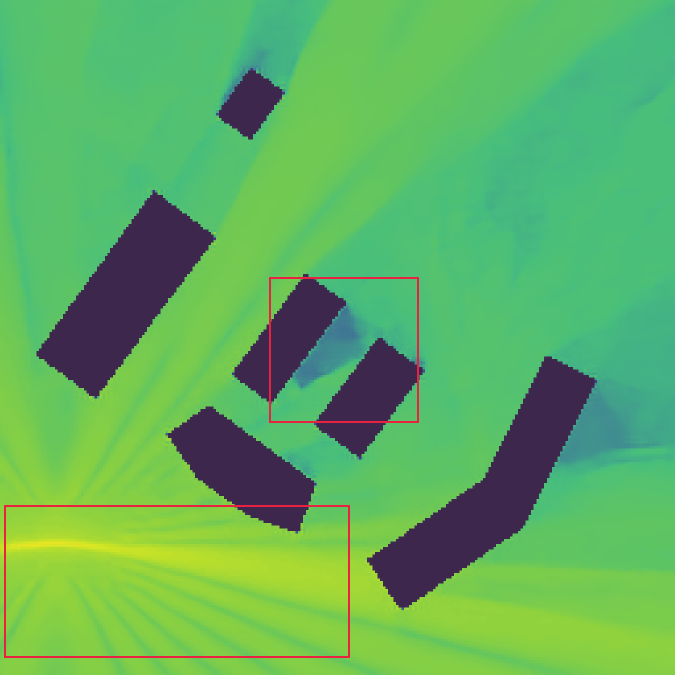}}
		\hspace{0.25cm}
		\subfloat{\includegraphics[width=1.1in]{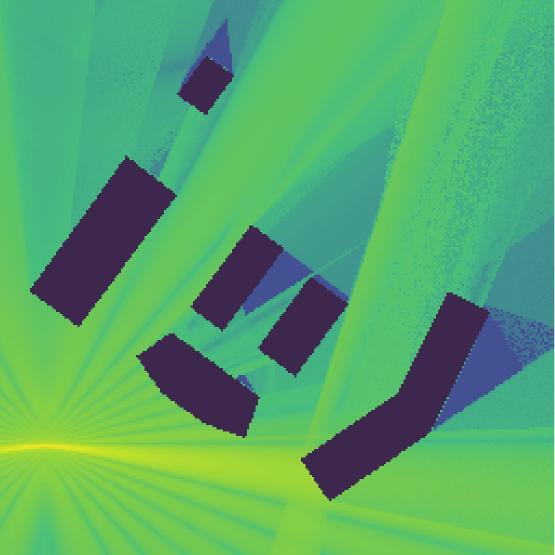}}\\
		\setcounter{subfigure}{0}
		\subfloat[RadioUNet]{\includegraphics[width=1.1in]{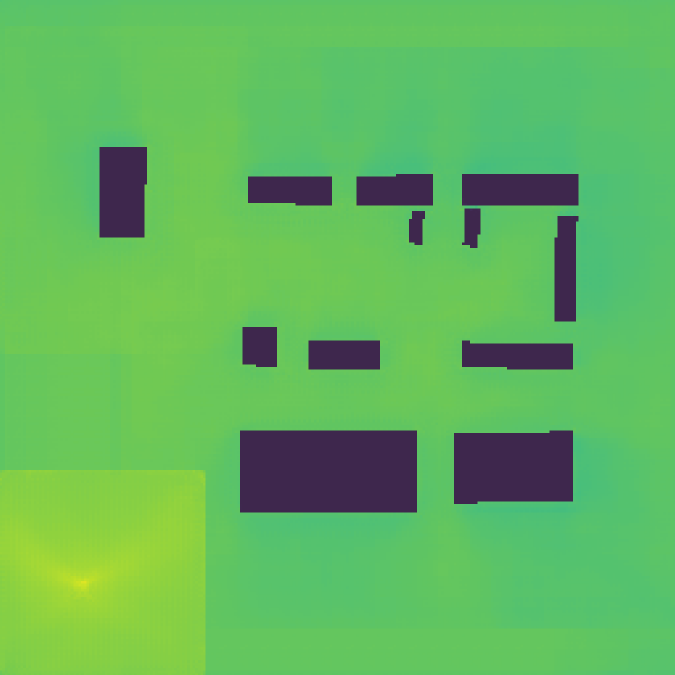}}
		\hspace{0.25cm}
		\subfloat[TransUNet]{\includegraphics[width=1.1in]{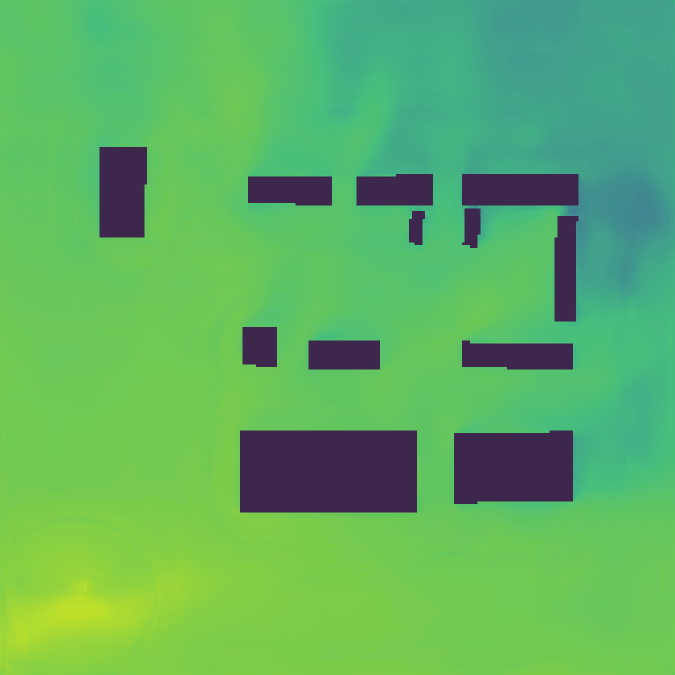}}
		\hspace{0.25cm}
		\subfloat[RadioDiff-UNet]{\includegraphics[width=1.1in]{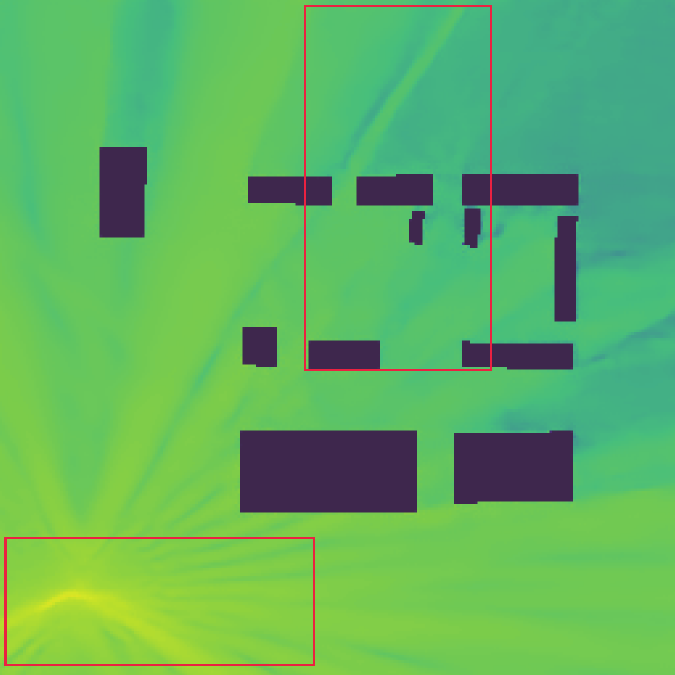}}
		\hspace{0.25cm}
		\subfloat[BeamCKMDiff]{\includegraphics[width=1.1in]{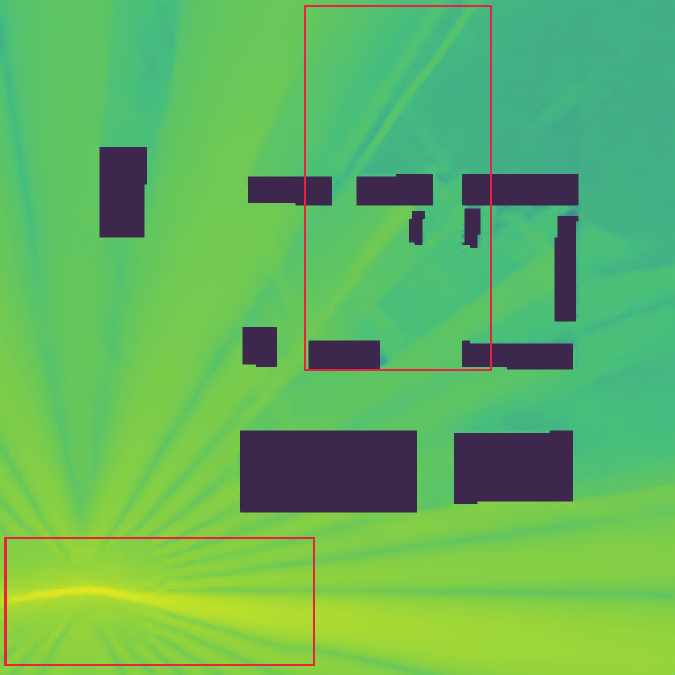}}
		\hspace{0.25cm}
		\subfloat[Ground Truth]{\includegraphics[width=1.1in]{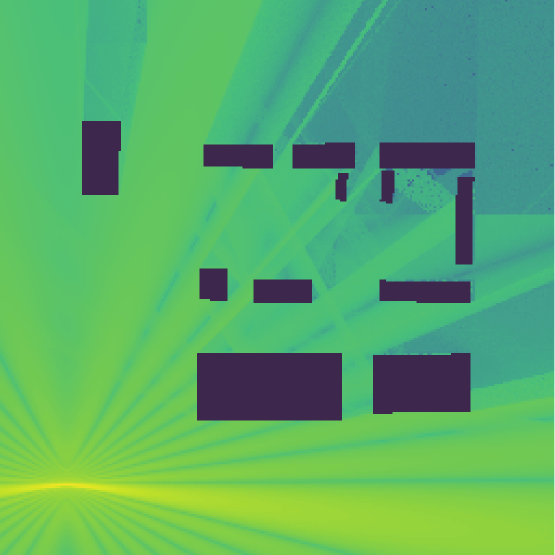}}
		\caption{Comparisons of constructed CKM via different methods under \textbf{unseen GBS locations and beams}.}
		\label{fig:visualization_umseemed_location}
		%		\vspace{-0.4cm}
	\end{figure*}
	
	\subsection{Results Discussion}
	
	Table \ref{tab:NMSE_performance_no_sample} presents the quantitative comparison. The proposed BeamCKMDiff achieves a remarkable NMSE lower than -20 dB, demonstrating superior reconstruction quality even without sampling measurements. This performance gain stems from the continuous beamforming vector conditioning, which provides explicit physical guidance for directional gain estimation. {Regarding computational efficiency, BeamCKMDiff records an average inference latency of 0.688 s. While the iterative diffusion process naturally incurs higher latency than discriminative baselines, the sub-second inference time remains practical for dynamic CKM construction, representing a justifiable trade-off for the significant accuracy enhancement.}
	\vspace*{-0.25cm}
	
	\begin{table}[!h]
		\renewcommand{\arraystretch}{1.1}
		\centering
		\caption{NMSE Performance Comparison}
		\label{tab:NMSE_performance_no_sample}
		\setlength{\tabcolsep}{1mm}
		\begin{tabular}{l|c|c|c}
			\hline
			\textbf{Methods} & \textbf{Unseen beams} & \textbf{Unseen locations} & \textbf{Inference time}\\
			\hline
			RadioUNet            & -16.35 dB & -16.34 dB & \textbf{0.231} s\\
			TransUNet       & -18.91 dB & -17.34 dB & 0.234 s\\
			RadioDiffUNet   & -19.49 dB & -19.13 dB & 0.373 s\\
			Beam-CKMDiff & \textbf{-21.24 dB} & \textbf{-20.68 dB} & 0.688 s\\
			\hline
		\end{tabular}
	\end{table}
	
	Visual comparisons in \myreffig{fig:visualization_umseemed_beam} and \myreffig{fig:visualization_umseemed_location} reveal distinct behavioral differences among methods. Lacking beam awareness, RadioUNet and TransUNet fail to reconstruct specific propagation paths, capturing only generic environmental shadowing. RadioDiff generates partial gain patterns but frequently misaligns the main lobe direction due to the absence of explicit beam guidance. In contrast, BeamCKMDiff precisely reconstructs both the directional beam patterns and fine-grained reflections/diffractions propagation effects.
	
	In the challenging unseen location scenario, while all models experience degradation in estimating environmental propagation interactions, BeamCKMDiff exhibits robust generalization by maintaining accurate beam pattern structures. Conversely, baseline methods suffer from severe beam misalignment. This confirms that the proposed beamforming-informed conditioning effectively disentangles beam steerability from spatial geometry, ensuring robust generalization across novel configurations.

	%	\vspace*{-0.2cm}
	
	\section{Conclusion}
	\label{sec:conclusion}
	In this paper, we proposed BeamCKMDiff, a generative framework for constructing beam-aware CKMs. By replacing the standard U-Net backbone with a DiT and incorporating a beamforming-informed adaLN mechanism, our model successfully decouples beam steering information from environmental geometries. {Simulation results demonstrated its construction capability within the continuous beam manifold, achieving an NMSE below -20 dB in a strictly sampling-free setting and outperforming state-of-the-art baselines. As the first work on beam-aware CKM construction, BeamCKMDiff sets a new benchmark for future research. In future work, we will focus on how to refine BeamCKM with sparse RSS measurements and investigate the optimal RSS sampling ratio to achieve an accuracy-cost trade-off.}
	
	%	\vspace*{-0.03cm}
	% BibTeX references generated from the provided PDF contexts
	\bibliographystyle{IEEEtran}
	\bibliography{IEEEabrv, Ref} % Ref.bib
	
\end{document}